\documentstyle[11pt]{article}
\font\tenbf=cmbx10
\font\tenrm=cmr10
\font\tenit=cmti10
\font\elevenbf=cmbx10 scaled\magstep 1
\font\elevenrm=cmr10 scaled\magstep 1
\font\elevenit=cmti10 scaled\magstep 1

\renewenvironment{thebibliography}[1]
 { \elevenrm
   \begin{list}{\arabic{enumi}.}
    {\usecounter{enumi} \setlength{\parsep}{0pt}
     \setlength{\itemsep}{3pt} \settowidth{\labelwidth}{#1.}
     \sloppy
    }}{\end{list}}

\begin{document}
\begin{flushright}
FSU--HEP--930923\\
September 1993
\end{flushright}
\begin{center}{
\vglue 0.6cm
{\elevenbf ELECTROWEAK PHYSICS AT CURRENT ACCELERATORS AND THE SUPERCOLLIDER
\footnote{To appear in the Proceedings of the Workshop {\it ``Physics at
Current Accelerators and the Supercollider''}, Argonne National
Laboratory, June~2 --~5, 1993.}
\\}
\vglue 1.0cm
{\tenrm presented by\\}
\vglue 0.3cm
{\tenrm U. BAUR
\footnote{Research supported by the
U.~S.~Department of Energy under Contract No.~DE-FG05-87ER40319.}
\\}
\baselineskip=13pt
{\tenit Physics Department \\}
\baselineskip=12pt
{\tenit Florida State University, Tallahassee, FL 32306\\}
\vglue 1.cm
{\tenbf Working Group Members}
\vglue 0.4cm
\baselineskip=13pt
{\tenrm U. BAUR$^1$\footnote{Co-convener}, S. ERREDE$^{2\ddagger}$, F.
HALZEN$^3$, S. KELLER$^1$, G.
LANDSBERG$^4$, M. L. MANGANO$^5$, A. PERYSHKIN$^6$, K. RIESSELMANN$^3$,
A. STANGE$^7$, and D. ZEPPENFELD$^3$\\}
\vglue 0.3cm
\baselineskip=12pt
{\tenit $^1$Florida State University, $^2$University of Illinois, Urbana
-- Champaign, $^3$University of Wisconsin, Madison, $^4$SUNY at Stony
Brook, $^5$INFN, Pisa, Italy, $^6$Fermilab, $^7$Brookhaven National
Laboratory\\}
\vglue 0.8cm
{\tenrm ABSTRACT}}

\end{center}

\vglue 0.1cm
{\rightskip=3pc
 \leftskip=3pc
 \tenrm\baselineskip=12pt
 \noindent
The activities of the Electroweak Physics Working Group are summarized.
Three main issues are addressed: 1) prospects for measuring the $W$ mass
at the Tevatron using the inclusive electron energy spectrum, 2)
constraining the strange quark distribution function in $W$ + charm
production, and 3) possibilities to determine the three vector boson
couplings at the Tevatron and SSC.
\vglue 0.6cm}
{\elevenbf\noindent 1. Introduction}
\vglue 0.2cm
\baselineskip=14pt
\elevenrm
The study of electroweak processes is one of the main tasks of
experiments at current and future accelerators. In order to test the
Standard Model (SM) of electroweak interactions its parameters have to
measured as precisely as possible and compared to the SM prediction.
Instead of attempting a global analysis of the capabilities of present
and future collider experiments, the electroweak working group has
focused on the $W$ boson mass and the three vector boson couplings as
important parameters to be measured. In addition, the working group has
also investigated the prospects of constraining the strange quark
distribution function in $W$ plus charm quark production at the
Tevatron. The studies reported here are by no means complete. In most
cases there is substantial room for improvement left.

The mass of the $Z$ boson is presently known with a precision of about
7~MeV from experiments at LEP~\cite{MZ}. On the other hand, the uncertainty on
the $W$ mass from the combined UA2 and CDF data~\cite{PDG} is
approximately 270~MeV. Both, UA2 and CDF, used the transverse mass
$m_T(\ell,\nu)$, $\ell=e,\,\mu$ of the charged lepton neutrino system in
$W\rightarrow\ell\nu$ decays in the past to determine $m_W$, a procedure
which explicitly depends on measuring the missing transverse
momentum, $p\llap/_T$, in the event. A significant fraction of the
overall uncertainty in the $W$ mass, $m_W$, originates from
systematic uncertainties associated with the $p\llap/_T$ measurement. A
quantity which is sensitive to $m_W$ but avoids the experimental
complications of the $p\llap/_T$ measurement is the lepton energy
spectrum. The prospects of using the electron energy distribution as an
alternative to the transverse mass for a high precision measurement of
$m_W$ were thoroughly investigated~\cite{AP}. The results of this
analysis are described in Section~2.

With the integrated luminosity accumulated during the 1992~--~94
Tevatron collider runs,
CDF and D\O\ expect several thousand $W+1$~jet events each.
For such a large $W+1$~jet sample it may become possible to
search for associated $W$ plus charm quark production. To lowest order
$Wc$ production proceeds via the fusion of a gluon and a $s$ or $\bar
s$-quark. A measurement of the $W$ plus charm production cross section
may thus help in resolving the controversy~\cite{WU} between the
MRS~\cite{MRS} and CTEQ~\cite{CTEQ}
parametrization of the strange quark distribution function. The
results of the $W+$~charm analysis are reported in Section~3~\cite{BHKMR}.

Within the SM, at tree level, the vector boson self-interactions are
completely fixed
by the $SU(2)\times U(1)$ gauge theory structure of the model. Measuring
the $WW\gamma$, $ZZ\gamma$, $Z\gamma\gamma$ and $WWZ$ couplings
therefore is a crucial test of the SM. At the Tevatron, this can be
accomplished studying $W^\pm\gamma$ and $Z\gamma$ production. Since the
number of events expected in run 1a+1b is quite limited, it is
important to optimize
the statistical procedure used to extract information on the three
vector boson vertices. Based on the maximum likelihood technique, a new
fitting procedure has been developed~\cite{GREG}, which considerably
improves upon the simple $\chi^2$ test used in most theoretical
simulations~\cite{BZ,NLOTWO}. The new technique is summarized in Section~4.1.

A pronounced feature of $W\gamma$ production in hadronic collisions is
the so-called radiation zero which appears in the parton level
subprocesses which contribute to lowest order in the SM of
electroweak interactions~\cite{RAZ}. In practice, however, this zero
is difficult to
observe. The quantity which represents the radiation zero best and, at
the same time, is easily measured experimentally is the distribution of
the photon -- lepton rapidity difference. This quantity was studied
extensively in the context of the Workshop~\cite{BEL} (see Section~4.2).

At the Tevatron, the $WWV$, $V=\gamma,\, Z$ and $ZZ\gamma$
($Z\gamma\gamma$) couplings cannot be measured very precisely, due to
the limited statistics of di-boson events. High precision tests have to
await the SSC or LHC~\cite{BZ,NLOTWO}. In order to probe the interactions in
the bosonic sector of the SM as accurately as
possible, all the important background processes need to be controlled.
Furthermore, one would like to have available a large
number of basic processes and correspondingly a large number of
observables. A potentially dangerous background to $W^\pm\gamma$
production at hadron supercolliders is $t\bar t\gamma$ production. The
working group found that the
$pp\rightarrow t\bar t\gamma+X$ cross section is much larger than the
lowest order $W\gamma$ rate; however, imposing a jet veto
requirement this background can be eliminated rather easily~\cite{BS}
(see Section~4.3). Besides the classic di-boson production processes
$q\bar q'\rightarrow W\gamma,\, WZ$, single $W$ production via the
electroweak process $qq\rightarrow qqW$ was studied as a complementary
source of information on the three vector boson couplings~\cite{UD}.
A description of the results obtained for the $qq\rightarrow qqW$
signal and the most
important background processes can be found in Section~4.4.
\vglue 0.3cm
{\elevenbf\noindent 2. Measuring the $W$ Boson Mass using the Inclusive
Electron Energy Spectrum}
\vglue 0.2cm
The SM has been extraordinarily successful in describing weak and
electromagnetic phenomena over the full reach of experimental
observation. The model is completely determined by three parameters: the
fine structure constant, $\alpha$, the Fermi constant, $G_F$, and the
$Z$ boson mass, $m_Z$. All three are known with high precision. With the
model so fixed, all other parameters, in particular the $W$ boson mass,
$m_W$, are determined. It is now the task of experiment to carry out
precision measurements of $m_W$ to confront the SM prediction.

At hadron colliders, the major problem of the $W$ mass measurement is
the neutrino originating from the $W\rightarrow\ell\nu$ decay which
escapes undetected. Thus there is no direct way to reconstruct $m_W$. A
standard way to overcome this problem is to use the transverse mass
distribution which sharply peaks at $m_W$. A limiting factor in the
determination of $m_W$ from the transverse mass distribution is the systematic
uncertainty associated with the determination of the missing transverse
momentum, $p\llap/_T$. Uncertainties originating from the $p\llap/_T$
measurement typically are at least
twice as large those from the measurement of the electromagnetic
energy.

As an alternative to the transverse mass spectrum, one can try to
utilize the lepton energy distribution, which sharply peaks at $m_W/2$
(see Fig.~1 of Ref.~3), and which requires information on the lepton
four-momentum only. Two different strategies of extracting $m_W$ from
the lepton energy distribution were studied. For definiteness, only
the decay $W\rightarrow e\nu$ was considered. In the first approach the
energy spectrum is fitted to an analytical function of the form
\begin{equation}
N_e(E)={N_0\over\Gamma+(E-m_W/2)^2}
\end{equation}
where $N_0$, $\Gamma$ and $m_W$ are free parameters of the fit which
correspond to the number of events, the width and the position of the
peak.

In the second approach, a direct comparison of the Monte Carlo electron
energy spectrum with the data is carried out, using a Kolmogorov test to
compare the two distributions. To estimate the statistical and
systematic uncertainties of this method, $10^5$ ISAJET $p\bar p\rightarrow
W\rightarrow e\nu$ events were generated, imposing a $E_t^e >25$~GeV, a
$E\llap/_t>25$~GeV, and a $|\eta_e|<1.1$ cut. To simulate detector
response the electron momentum four vector was smeared with a Gaussian
distribution with standard deviation $\sigma=0.15~{\rm GeV}^{1/2}\sqrt{E}$.
The MC event sample was split into a ``basic'' sample of 70,000 events,
and a ``test'' sample which varied from 1,000 to 30,000 events. The
statistical uncertainty as a function of the number of events in the test
sample is shown in Fig.~1. The solid line corresponds to a fit of the
uncertainty of the form $P/\sqrt{N_{ev}}$ where $P$ is a free fit parameter.
\begin{figure}[t]
\vskip 2.2in
\includegraphics{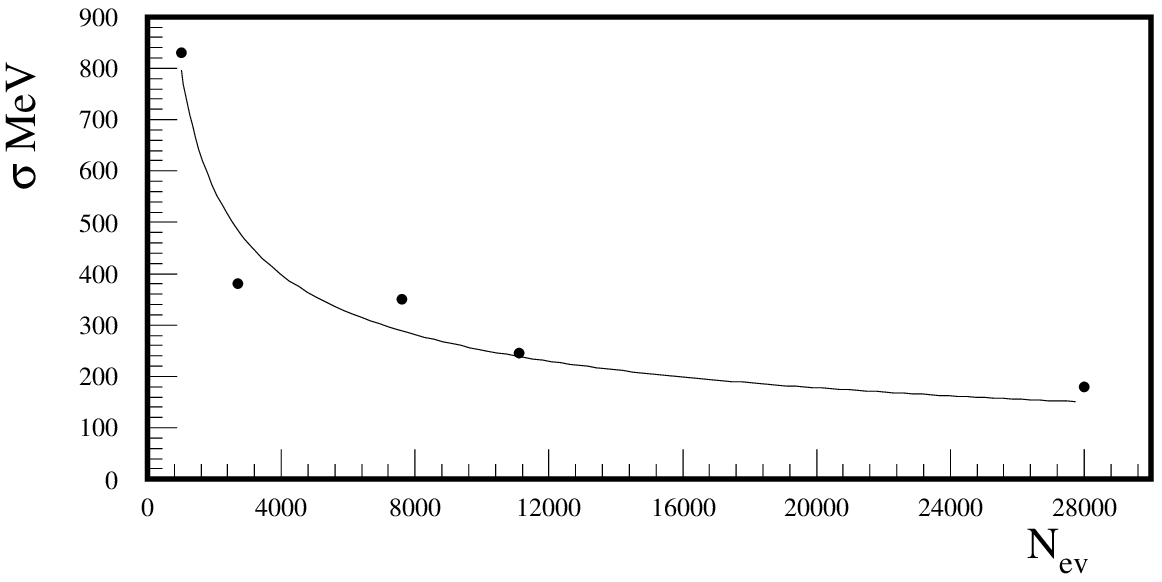}
Figure~1: The statistical uncertainty versus number of events, $N_{ev}$, in
the test sample. The solid line shows the result of fitting the uncertainty
with the function $P/\sqrt{N_{ev}}$.
\end{figure}
For 40,000 $W$ events, which roughly corresponds to the number of
$W\rightarrow e\nu$ events
anticipated at the Tevatron for an integrated luminosity of
100~pb$^{-1}$, the expected statistical uncertainty is about 125~MeV,
and thus about twice as large as the uncertainty from the conventional method
employing the transverse mass distribution.

The systematic uncertainties originating from the transverse motion of the $W$,
and from varying the cuts imposed on $E\llap/_t$,
$E_t^e$, and $\eta_e$ were found to be in
the 50~--~100~MeV range each. The expected combined systematic uncertainty is
approximately 150~MeV, and thus significantly smaller than that of the
conventional method utilizing the transverse mass distribution.
\vglue 0.3cm
{\elevenbf\noindent 3. The Charm Content of $W+1$~Jet Events and the
Strange Quark Distribution Function}
\vglue 0.2cm
Approximately 15\% of the inclusive number of $W$ bosons produced at the
Tevatron are accompanied by a hadronic jet with a transverse energy
$E_t^{jet}>15$~GeV \cite{CDF}. From run~1a and~1b combined one therefore
expects more than 6,000 $W+1$~jet events with $W\rightarrow e\nu$. The
large number of $W$ events accompanied by a high $p_T$ jet will make it
possible to specifically search for heavy quarks in such events. In
particular events for which the jet contains a charm quark could be
useful, as they may allow to constrain the strange quark distribution
function.

Recent fits of parton distribution functions by the MRS~\cite{MRS} and
CTEQ collaborations~\cite{CTEQ} have resulted in rather different
$s$-quark distribution functions. The ratio of the strange quark
distribution functions of the CTEQ1M and the MRSD0 sets versus $x_s$
is shown in Fig.~2a.
\begin{figure}[t]
\vskip 2.8in
\includegraphics{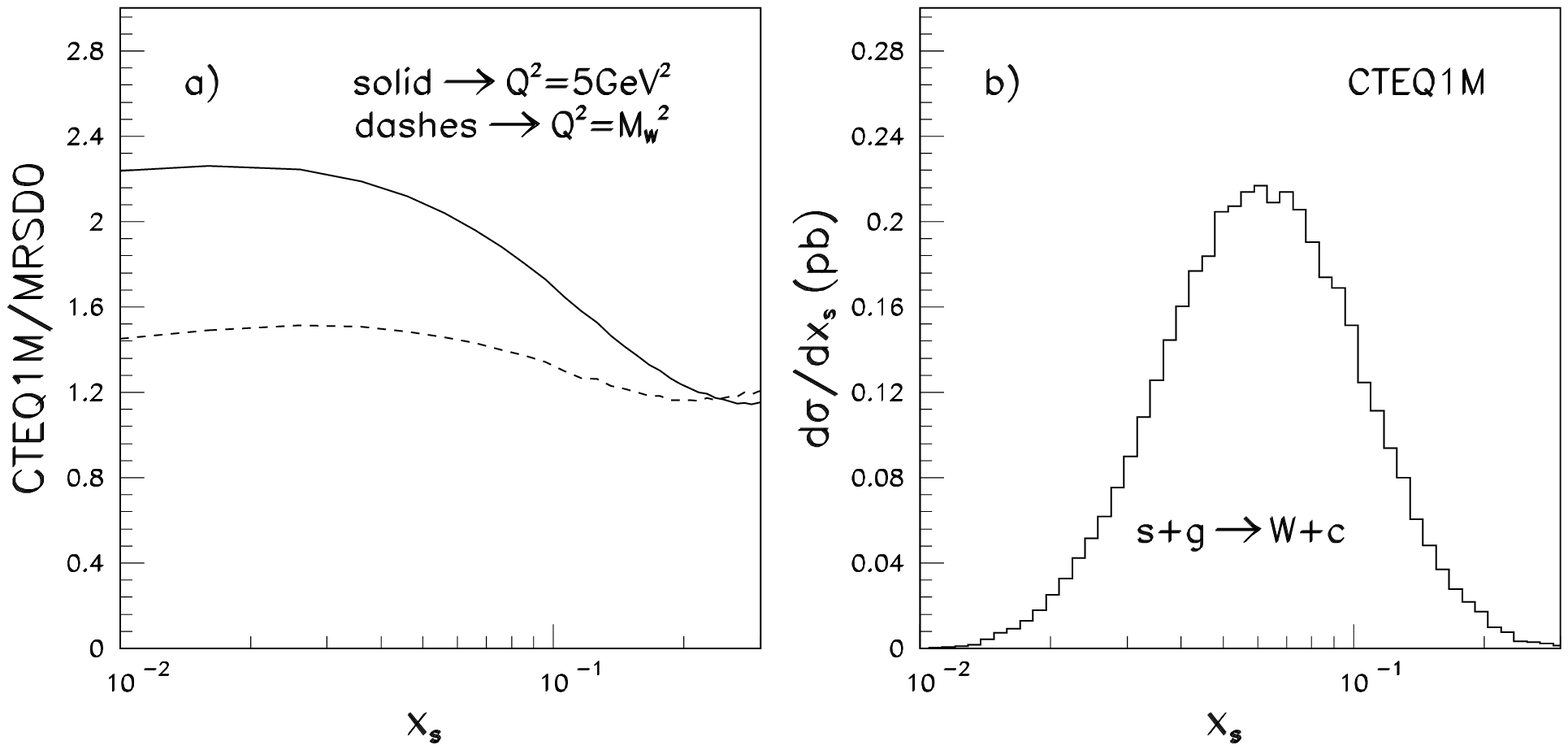}
Figure~2: a) Ratio of the strange quark distribution functions as a
function of $x_s$ for the CTEQ1M and MRSD0 parton distribution sets
and two different values of $Q^2$.
b) The differential cross section $d\sigma/dx_s$ for the process
$sg\rightarrow Wc$ at the Tevatron, using the CTEQ1M set.
\end{figure}
At small $Q^2$ and $x_s < 0.1$, the two parametrizations
differ approximately by a factor of two.

If the charm content of $W+1$~jet events can be determined, an
independent measurement of the strange quark
distribution function can be carried out. Associated $W+$~charm
production proceeds, at lowest order, through $sg$
and $\bar sg$ fusion, $sg\rightarrow W^-c$ and $\bar sg\rightarrow
W^+\bar c$. The alternative process where the $s$-quark in the reaction is
replaced by a $d$-quark, is suppressed by the quark mixing matrix
element $V_{cd}$. This suppression is somewhat compensated by the
larger $d$ quark
distribution function, such that the $dg\rightarrow Wc$ cross section is
about 10\% of the $sg\rightarrow Wc$ rate. The potentially largest
background originates from the production of a
$c\bar c$ pair in the jet recoiling against the $W$.  When only the $c$
or the $\bar c$ is identified in
the jet, such a $W+c\bar c$ event looks like a signal event.
Similarly, a $b\bar b$ pair can be produced in the jet, and the $b$ or the
$\bar b$-quark misidentified as a charm quark.

To numerically simulate the signal and background processes
the Monte--Carlo program PYTHIA~\cite{PYT87} (version~5.6) was used.
All processes were studied at the parton level, {\it i.e.} final state
showers are included but fragmentation is not. In the
simulations carried out,
a ``jet'' is defined as follows. The direction of the sum of the momenta
of all the partons produced in the shower is taken as the center of a cone
of radius $\Delta R_{jet}=\sqrt{\Delta\eta^2 +\Delta\phi^2} = 0.7$,
where $\eta$ is the pseudorapidity
and $\phi$ the azimuthal angle. All the partons inside that cone are
considered part of the jet.  Only events with a charm quark {\it inside}
the jet cone are counted. Background
events with two charm quarks inside the jet cone are counted twice.

The $W^\pm$ is assumed to decay into a $e^\pm\nu$ final state. To simulate
the acceptance of a real detector,
the following transverse momentum and pseudorapidity cuts on the
final state particles are imposed:
\begin{eqnarray}
\label{eq:cut}
p_{T}(e)  & \geq & 20~{\rm GeV}, \hskip 1.cm |\eta(e)| \leq 1,  \nonumber
\\
p\llap/_{T} & \geq & 20~{\rm GeV}, \hskip 1.cm \phantom{|\eta(e)| \leq 1.}
 \\
p_{T}(j) & \geq & 10~{\rm GeV}, \hskip 1.cm |\eta(j)| \leq 1. \nonumber
\end{eqnarray}
Figure~2b shows the differential
cross section of the process $sg\rightarrow Wc$ for $p\bar p$
collisions at $\sqrt{s}=1.8$~TeV as a function of the momentum fraction of
the strange quark, $x_s$, using the CTEQ1M parametrization.
$W$ plus charm quark production at the
Tevatron thus is sensitive to the strange quark distribution mostly in
the $x_s$ region between 0.04 and 0.1, in which the CTEQ and MRS
parametrizations are indeed substantially different (see Fig.~2a).

The cross sections for the signal, the various background processes, and
inclusive $W+1$~jet production at the Tevatron are given in
Table~1.  Approximately 75\% (20\%) of the
background originates from a $c\bar c$ ($b \bar b$) pair produced in a jet
initiated by a gluon, if all $b$ and $\bar b$-quarks are assumed to be
misidentified as charm quarks.
The remaining 5\% is due to the production of a $c \bar c$ pair in a
quark--initiated jet. The combined background cross section is about equal
to the signal rate. The signal accounts for approximately 4--5\% of
all $W+1$~jet events, and the background for about 4\%.
As can be seen from Table~1, the two sets of parton distribution
functions yield the same values for the inclusive $W+1$~jet cross section
and the three background processes to within 2\%.
The signal rate, on the other hand, is quite sensitive to which set is
chosen, as expected.
\begin{table}[t]
\centering
\caption
[Cross sections.]
{Cross sections for associated $W$ plus charm quark production with
$W\rightarrow e\nu$ at the Tevatron, using the MRSD0 and CTEQ1M
parametrizations of the parton distribution functions. The cuts imposed
are summarized in Eq.~2.}
\vspace{6.mm}
\begin{tabular}{|c||c|c|c|c|c|}\hline
        & \multicolumn{5}{c|}{Cross section (pb)} \\ \cline{2-6}
	& $Wc$ & \multicolumn{3}{c|}{$W+1$~jet background} & Inclusive \\
\cline{3-5}
	& Signal & $g\rightarrow c\bar c + X$
&$g\rightarrow b\bar b + X$ &$q\rightarrow qc\bar c + X $& $W$+ 1 jet \\
\hline\hline
MRSD0  & 3.68 & 2.76 & 0.75 & 0.15 & 87.0 \\ \hline
CTEQ1M & 4.58 & 2.80 & 0.77 & 0.15 & 85.6 \\ \hline
\end{tabular}
\end{table}

Assuming both electron and muon decay channel of the $W^{\pm}$ boson,
an integrated luminosity of 10~pb$^{-1}$ yields about 1700
$W+1$~jet events for the cuts described in Eq.~2. This corresponds to
approximately 75 -- 90 $W$ plus charm quark signal events, and to about
the same number of potential background events. From these numbers it is
straightforward to estimate the minimum charm tagging efficiency,
$\epsilon_c^{min}$, required to be statistically sensitive to the
variation of the $Wc$ production cross section with the strange quark
distribution function. Depending on how efficiently the various
background processes can be
suppressed, an efficiency of $\epsilon^{min}_c\approx 20 - 30$\% is
needed for an integrated luminosity of 10~pb$^{-1}$.

The two collider experiments, CDF and D\O, at the Tevatron explore
three different strategies to identify charm quarks:
\begin{enumerate}
\item Search for a displaced secondary vertex in the silicon vertex
detector (SVX).
\item Reconstruction of exclusive nonleptonic charmed baryon or
meson decays.
\item Looking for inclusive semileptonic charm decays.
\end{enumerate}

Combined, the three methods may yield an overall charm detection
efficiency of about 10\%. Based on this assumption, an integrated
luminosity
of ${\cal O}(30$~pb$^{-1})$ should provide the first statistically
significant information on the strange quark distribution of the proton.
A more precise estimate of the minimum integrated luminosity required
depends on a better understanding of the charm quark detection
efficiency, and on more detailed background studies. In
principle, the three background processes considered here can be
reduced by:

\begin{itemize}
\item Charge reconstruction: for the signal, the $W$ and $c$-quark
electric charges are correlated. For the $c\bar c$ background, the
charm quark has the wrong charge 50\% of the time. Therefore, if the
charges of the $W$ and of the charm quark can be determined, the $Wc\bar c$
background can be reduced by
a factor of two.  Furthermore, events with the wrong charge correlation
provide a measurement of the background, that could subsequently be
subtracted.
\item Cut on the charm transverse momentum: since more than
one charm quark is present in the background processes its average $p_T$
is smaller than in the signal.
\item Flavor identification: if the bottom quark is
identified, the $b\bar b$ background can be subtracted.
\end{itemize}

Clearly, more experimental and theoretical work is needed to improve
the signal to background ratio.
\vglue 0.3cm
{\elevenbf\noindent 4. Probing the Vector Boson Self-interactions in Hadron
Collider Experiments}
\vglue 0.2cm
{\elevenit\noindent 4.1 Extracting Three Vector Boson Couplings from the
Photon Transverse Momentum Distribution}
\vglue 0.1cm
The increased integrated luminosity accumulated in the 1992~--~94 Tevatron
collider runs will not only make it possible to improve the precision of
existing measurements, but also to probe previously untested sectors of
the SM, such as the self-interactions of vector bosons. This is most
easily done studying $W\gamma$ and $Z\gamma$ production.

The most general $WW\gamma$ vertex function for $q\bar q'\rightarrow
W\gamma$ which is compatible with electromagnetic gauge invariance and
Lorentz invariance, and which is $CP$ conserving, can be parameterized in
terms of two couplings $\kappa^\gamma$ and $\lambda^\gamma$~\cite{BZ}.
Similarly, under the same conditions, the
$ZZ\gamma$ and $Z\gamma\gamma$ vertex can be described by two couplings,
$h^V_3$ and $h^V_4$, $V=\gamma,\, Z$. In the SM, at tree level,
$\kappa^\gamma=1$, $\lambda^\gamma=0$, and $h^V_3=h^V_4=0$. In order to
avoid violations of
$S$-matrix unitarity, the anomalous three vector boson couplings have to
be introduced as momentum dependent form factors. Frequently, generalized
dipole form factors of the form
\begin{equation}
a(\hat s) = {a_0\over (1+\hat s/\Lambda^2)^n}~,
\end{equation}
with $a=\kappa^\gamma-1,\dots,h^V_4$ are used. $\Lambda$ in Eq.~(3) represents
the scale at which new physics becomes important in the weak boson
sector, $\hat s$ is the energy squared in the parton center of mass
frame, and $a_0$ are the form factors at low energy. In order to
guarantee unitarity, $n$ must satisfy $n>1/2$ for
$\Delta\kappa^\gamma=\kappa^\gamma-1$, $n>1$ for $\lambda^\gamma$,
$n>3/2$ for $h^V_3$, and $n>5/2$ for $h^V_4$.

Non-standard $WW\gamma$ and $ZZ\gamma$ ($Z\gamma\gamma$) couplings lead
to a broad increase in the photon transverse momentum distribution at
large values of the photon $p_T$~\cite{BZ,NLOTWO}. Information on the anomalous
three vector boson couplings thus can be derived by fitting the observed
$p_T(\gamma)$ distribution with general $WW\gamma$ and $ZZ\gamma$
($Z\gamma\gamma$) couplings.

The main background for $p\bar p\rightarrow W\gamma$, $Z\gamma$ is
$W+jets$ and $Z+jets$ production
with one of the jets faking a high $p_T$ photon due to fragmentation of
the jet into a high energy $\pi^0$ or $\eta$, which decays into two
almost collinear photons which are not resolved in the detector. This
background rapidly drops with the $\gamma/jet$ transverse momentum. The
probability $P_{\gamma/j}$ for the jet faking a photon has been
measured by CDF and D\O\ separately. Both collaborations find
$P_{\gamma/j}\approx 10^{-3}$ for $p_T(\gamma)>10$~GeV. Electrons
misidentified as photons due to tracking inefficiencies may also
constitute a non-negligible background for D\O.

In most theoretical simulations~\cite{BZ,NLOTWO}, a rather primitive procedure
for fitting the photon $p_T$ spectrum was used. In this procedure the
distribution is split into a number of bins of equal bin width plus one
additional bin which contains all events above a certain $p_T$
threshold. This threshold is chosen such that each bin contains typically
more than 5 events. In each bin, the Poisson statistics is then
approximated by a Gaussian distribution, and a minimum $\chi^2$ test
is performed.

A substantial improvement of this method can be achieved, especially for
low statistics, by two simple modifications:
\begin{enumerate}
\item Replace the Gaussian distribution and the $\chi^2$ test by Poisson
statistics and a maximum likelihood fit. In contrast to the
$\chi^2$ method, the maximum likelihood technique yields unbiased
estimates of the fitted parameters.
\item An extra bin, in which one would not expect any SM events at the
95\% confidence level (CL) is added. This modification fully exploits
the enhancement of the differential cross section at large $p_T(\gamma)$
values for anomalous couplings.
\end{enumerate}
The results of applying the improved fitting procedure to the data sets
expected for D\O\ are summarized in Table~2.
\begin{table}[t]
\centering
\caption
[limits.]
{Expected limits from D\O\ for the $CP$ conserving $WW\gamma$ and
$ZZ\gamma$ couplings. The limits for the $ZZ\gamma$ couplings are
derived using $n=3$ ($n=4$) for $h^Z_3$ ($h^Z_4$) and a from factor
scale of $\Lambda=750$~GeV. The bounds on the $WW\gamma$ couplings
$\Delta\kappa^\gamma$ and $\lambda^\gamma$ are insensitive to the form
factor details. Both, the electron and muon decay channels for the $W$
and $Z$ boson are used.}
\vspace{6.mm}
\baselineskip=15pt
\begin{tabular}{|c|c|c|c|c|}\hline
 Coup-  & \multicolumn{2}{c|}{run 1a limits} & \multicolumn{2}{c|}{run
 1a+1b limits} \\ \cline{2-5}
 ling   & 68\% CL & 95\% CL & 68\% CL & 95\% CL \\ \hline
$|\Delta\kappa^\gamma_0|$ & 1.10 & 2.50 & 0.60 & 1.15 \\
$|\lambda^\gamma_0|$ & 0.30 & 0.75 & 0.15 & 0.31 \\ \hline
$|h^Z_{30}|$ & 0.62 & 1.27 & 0.38 & 0.72 \\
$|h^Z_{40}|$ & 0.10 & 0.21 & 0.06 & 0.11 \\ \hline
\end{tabular}
\end{table}
They are up to a factor~2 better than those derived using the $\chi^2$
test. The limits for
$\Delta\kappa^\gamma_0$ ($h^Z_{3)}$) apply for arbitrary values of
$\lambda^\gamma_0$ ($h^Z_{40}$) and vice versa. Bounds on $h^\gamma_{30}$
($h^\gamma_{40}$) are approximately 10\% weaker than those for
$h^Z_{30}$~($h^Z_{40}$).
\vglue 0.2cm
{\elevenit\noindent 4.2 Rapidity Correlations in $W\gamma$ Production}
\vglue 0.1cm
A pronounced feature of $W\gamma$ production in hadronic collisions is
the so-called radiation zero which appears in the parton level
subprocesses which contribute to lowest order in the SM of
electroweak interactions~\cite{RAZ}. For $u\bar d\to W^+\gamma$ ($d\bar u\to
W^-\gamma$) all contributing
helicity amplitudes vanish for $\cos\Theta^*=-1/3$ (+1/3), where
$\Theta^*$ is the angle between the quark and the photon in the parton
center of mass frame. In practice, however, this zero is difficult to observe.
Structure function effects transform the zero into a dip. Higher order
QCD corrections \cite{NLOTWO,NLO} and finite $W$ width effects, together with
photon radiation from the final state lepton line, tend to fill in the
dip. Finally, the twofold ambiguity in the reconstructed parton center
of mass frame which originates from the two possible solutions for the
longitudinal momentum of the neutrino~\cite{STROUGHAIR}, $p_L(\nu)$,
represents
an additional complication in the extraction of the $\cos\Theta^*$ or
the corresponding rapidity distribution, $d\sigma/dy^*(\gamma)$, which further
dilutes the effect.

Photon lepton rapidity correlations provide an alternative tool to
search for the radiation zero predicted by the SM for $W\gamma$
production in hadronic collisions. For the remainder of this subsection
we shall focus entirely on the $W^+\gamma$ channel.
Results for $W^-\gamma$ production can be obtained by simply exchanging the
sign of the rapidities involved. The SM radiation zero leads to a
pronounced dip in the photon rapidity distribution in the center of
mass frame, $d\sigma/dy^*(\gamma)$, at
\begin{eqnarray}
y^*(\gamma)=y_0=-{1\over 2}\,\log 2\approx -0.35.
\end{eqnarray}
For $u\bar d\rightarrow W^+\gamma$ the photon and the $W$ are back to back
in the center of mass frame. The corresponding rapidity distribution of
the $W$ in the parton center of mass frame, $d\sigma/dy^*(W)$, thus
exhibits a dip at $y^*(W) = -y_0$. In the
double differential distribution of the rapidities in the laboratory
frame, $d^2\sigma/d\eta(\gamma)dy(W)$, one then expects a ``valley'' or
``channel'' for
rapidities satisfying the relation\footnote{Differences of rapidities are
invariant under boosts.} $\eta(\gamma)-y(W)\equiv y^*(\gamma)-y^*(W)=2y_0$.

In the SM, the dominant $W^\pm$ helicity in $W\gamma$ production is
$\lambda_W = \pm 1$~\cite{BBS}, implying that
the charged lepton will tend to be emitted in the direction of the parent
$W$, thus reflecting most of its kinematic properties. The difference
in rapidity, $\Delta y(W,\ell)=y(W)-\eta(\ell)$,
between the $W$ boson and the charged lepton originating from the $W$ decay
is rather small with an average $\Delta y(W,\ell)$ of 0.30.
The double differential distribution $d^2\sigma/d\eta(\gamma)d\eta(\ell)
$ for $p\bar p\rightarrow W^+\gamma\rightarrow\ell^+p\llap/_T\gamma$
thus displays a channel for rapidities fulfilling the relation
$\Delta\eta(\gamma,\ell)=\eta(\gamma) -\eta(\ell)\approx -0.4$.

The double differential distribution $d^2\sigma/d\eta(\gamma)d\eta(\ell)
$ can only be mapped out if a sufficiently large number of events is
available. For a relatively small event sample the distribution of the
rapidity difference, $d\sigma/d\Delta\eta(\gamma,\ell)$, which is shown
in Fig.~3, is more useful.
\begin{figure}[t]
\vskip 8. cm
\includegraphics{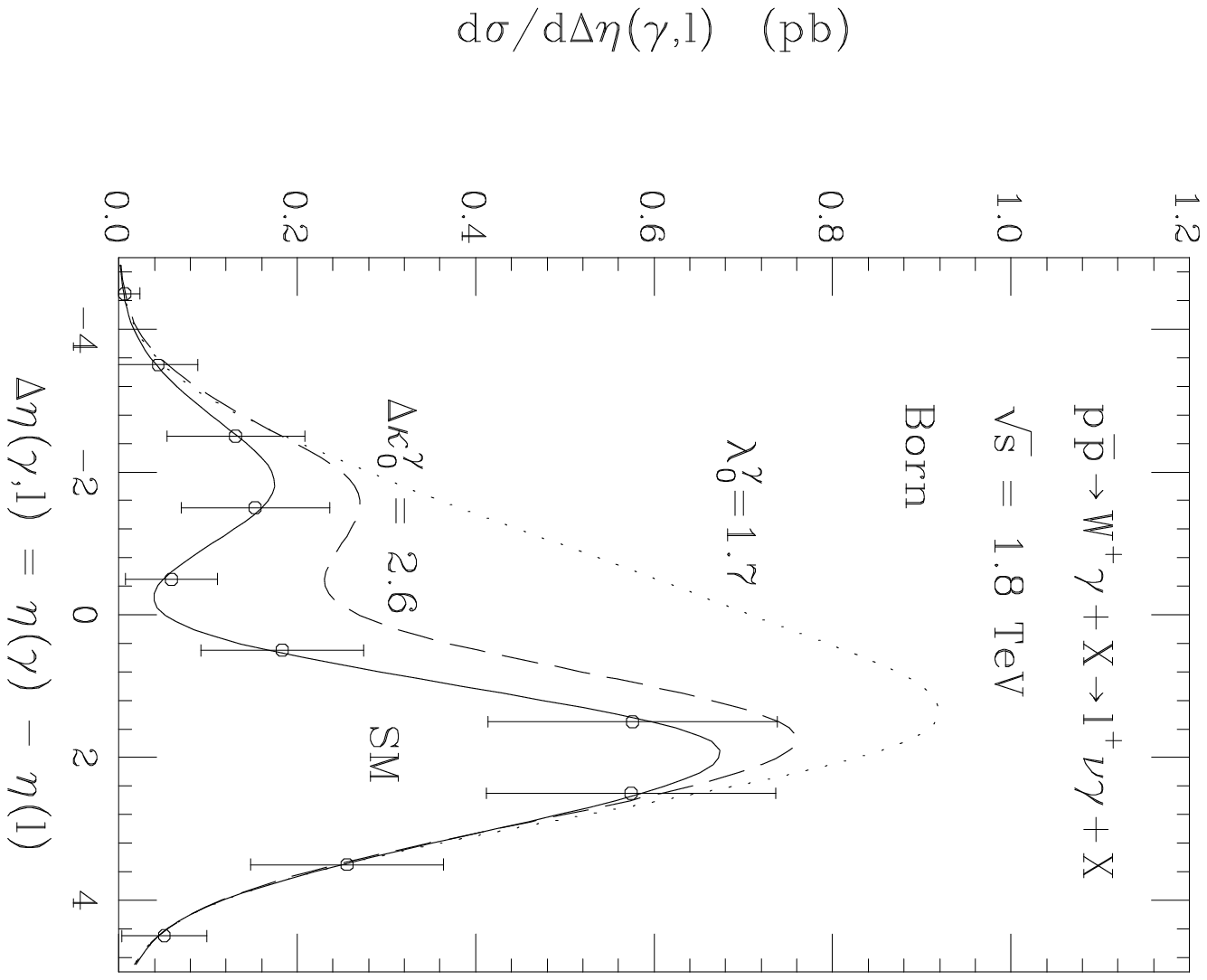}
\noindent Figure~3: The rapidity difference distribution,
$d\sigma/d\Delta\eta(\gamma,\ell)$, for $p\bar p\rightarrow
W^+\gamma+X\rightarrow\ell^+p\llap/_T\gamma+X$,  $\ell=e,\,\mu$, at
the Tevatron in the Born
approximation for anomalous $WW\gamma$ couplings. The curves are for
the SM (solid), $\Delta\kappa^\gamma_0=2.6$
(dashed), and $\lambda^\gamma_0=1.7$ (dotted). Only one coupling is
varied at a time. A dipole form factor with scale
$\Lambda=1$~TeV is assumed for non-standard $WW\gamma$ couplings. The
cuts imposed are described in the text. The error bars indicate the
expected statistical uncertainties for an integrated luminosity of
22~pb$^{-1}$.
\end{figure}
Here a $p_T(\gamma)>5$~GeV, a $p_T(\ell)>20$~GeV and a
$p\llap/_T>20$~GeV cut are imposed, together with cuts on the
pseudorapidities of
the photon and charged lepton of $|\eta(\gamma)|<3$ and $|\eta(\ell)| <
3.5$. To select a phase space region where radiative $W$ decays
are suppressed and $q\bar q'\rightarrow W\gamma$ dominates, we have
required in addition a large photon lepton
separation cut of $\Delta R(\gamma,\ell)>0.7$, and a cluster transverse
mass cut of $m_T(\ell\gamma;p\llap/_T) > 90$~GeV. The solid line shows
the SM result, whereas the dashed and
dotted curves give the prediction for the current UA2 68\% CL
limits~\cite{UA2} of $\Delta\kappa^\gamma_0=2.6$ and $\lambda^\gamma_0=1.7$,
respectively. In Fig.~3, a dipole form factor ($n=2$) with a scale
$\Lambda = 1$~TeV was used.

As anticipated, the $\Delta\eta(\gamma,\ell)$ distribution exhibits a
strong dip at $\Delta\eta(\gamma,\ell)\approx -0.4$ in the SM (solid
line). Higher order QCD corrections fill in the dip partially, but do
not obscure the radiation zero in a significant way. Compared to
$d\sigma/dy^*(\gamma)$, the rapidity difference distribution
has the advantage of being independent of the twofold ambiguity in the
reconstruction of the parton center of mass frame, which considerably
obscures the radiation zero in the $y^*(\gamma)$ distribution. In
contrast to the $Z\gamma$ to $W^\pm\gamma$ cross section ratio which
also reflects the radiation zero~\cite{BEO}, the rapidity difference
distribution
does not depend on the $Z\gamma$ cross section, and the validity of the
SM for $p\bar p\rightarrow Z\gamma$.

In presence of any anomalous contribution to the $WW\gamma$ vertex the
radiation zero is eliminated and the dip in $d\sigma/d\Delta\eta(\gamma,
\ell)$ is filled in at least partially. Most of the excess cross section
for non-standard couplings originates in the high $p_T(\gamma)$
region~\cite{BZ}, where events tend to be central in rapidity. Deviations from
the SM $\Delta\eta(\gamma,\ell)$ distribution, therefore, mostly occur
for small rapidity differences. In Fig.~3 we have also included the
statistical uncertainties expected in the SM case for an integrated luminosity
of $\int\!{\cal L} dt=22$~pb$^{-1}$. This demonstrates that the rapidity
difference distribution is sensitive to anomalous $WW\gamma$ couplings
already with the current CDF and D\O \ data samples, in particular
to $\lambda^\gamma$. However, one does not expect $d\sigma/d\Delta\eta(\gamma,
\ell)$ to be more sensitive to anomalous couplings than the photon
transverse momentum distribution.
\vglue 0.2cm
{\elevenit\noindent 4.3 The $t\bar t\gamma$ Background in $W\gamma$ Production
at the SSC}
\vglue 0.1cm
Whereas the three vector boson couplings can be measured only at the
10~--~60\% level at best at the Tevatron (see Section~4.1), the SSC or
LHC provide the energy and luminosity necessary for a high precision
test of these couplings. However, for $W\gamma$ production
the process $pp\rightarrow t\bar
t\gamma\rightarrow W\gamma+X$ represents a potentially dangerous
background, due to the very large top quark production cross section at
supercollider energies. In this subsection a calculation of
the $t\bar t\gamma$ background at the SSC is described. The calculation
fully incorporates
the subsequent decay of the top quarks into a $W$ boson and a $b$-quark,
and also the $W$ decay into a fermion antifermion pair. Graphs where the photon
is radiated from one of the $t$ or $\bar t$ decay products are not
taken into account. The contribution from these diagrams is strongly
suppressed if
a photon $p_T$ cut of $p_T(\gamma) > m_t/2$ is imposed. The top quark
and $W$ boson decays are treated in the narrow width approximation.

Besides the lowest order contributions to the associated production of a
$t\bar t$ pair and a photon, the photon bremsstrahlung contribution in
$t\bar tj$ events is included in our calculation. The bremsstrahlung
contribution is calculated using the QCD $q\bar
q\rightarrow t\bar tg$, $qg\rightarrow t\bar tq$ and $gg\rightarrow
t\bar tg$ matrix elements together with the leading-logarithm
parametrization of Ref.~\cite{DO} for the photon fragmentation
functions.
These fragmentation functions are proportional to $\alpha/\alpha_s$, and
the photon bremsstrahlung contribution formally is of the same order in
$\alpha$ as the lowest order $t\bar t\gamma$ cross section.

In the following only the channel $pp\rightarrow W^+\gamma+X$ is
considered. The cross sections of the $t\bar t\gamma$ background are equal
for the $W^+\gamma+X$ and $W^-\gamma+X$ channel. The $pp\rightarrow
W^-\gamma+X$ signal rate is approximately 20\% smaller than the $pp\rightarrow
W^+\gamma+X$ cross section for the cuts specified below. The conclusions
drawn for the $W^+\gamma$ case therefore also apply to the $W^-\gamma$
channel.

The $W^+$ boson is assumed to decay into a
$\ell^+\nu$ final state with $\ell=e,\,\mu$. In order to simulate the
finite acceptance of detectors and to reduce fake backgrounds from jets
misidentified as photons and particles lost in the beam
pipe~\cite{SDC}, the following transverse momentum, pseudorapidity and
separation cuts are imposed:
\begin{eqnarray}
p_T(\ell^+) > \phantom{1}25~{\rm GeV}, & \qquad & ~~~~|\eta(\ell^+)| < 3.0,\\
p_T(\gamma) > 100~{\rm GeV}, & \qquad & ~~~~~|\eta(\gamma)| < 2.5,\\
p\llap/_T > \phantom{1}50~{\rm GeV}, & \qquad & \Delta R(\ell^+,\gamma) > 0.7.
\end{eqnarray}
No cuts are imposed on the $b$-quark jets and the decay products of the
second $W$ ({\elevenit i.e.} the $W^-$) in $t\bar t\gamma$ events.
$W^-\rightarrow\tau\nu_\tau$ decays
are, for simplicity, treated like $W^-\rightarrow e\nu,\,\mu\nu$.

Figure~4a shows the $p_T(\gamma)$ distribution for $pp\rightarrow t\bar
t\gamma+X\rightarrow\ell^+p\llap/_T\gamma+X$ at the SSC for
$m_t=110$~GeV (solid line) and $m_t=200$~GeV (dashed line).
\begin{figure}[t]
\vskip 8.cm
\includegraphics{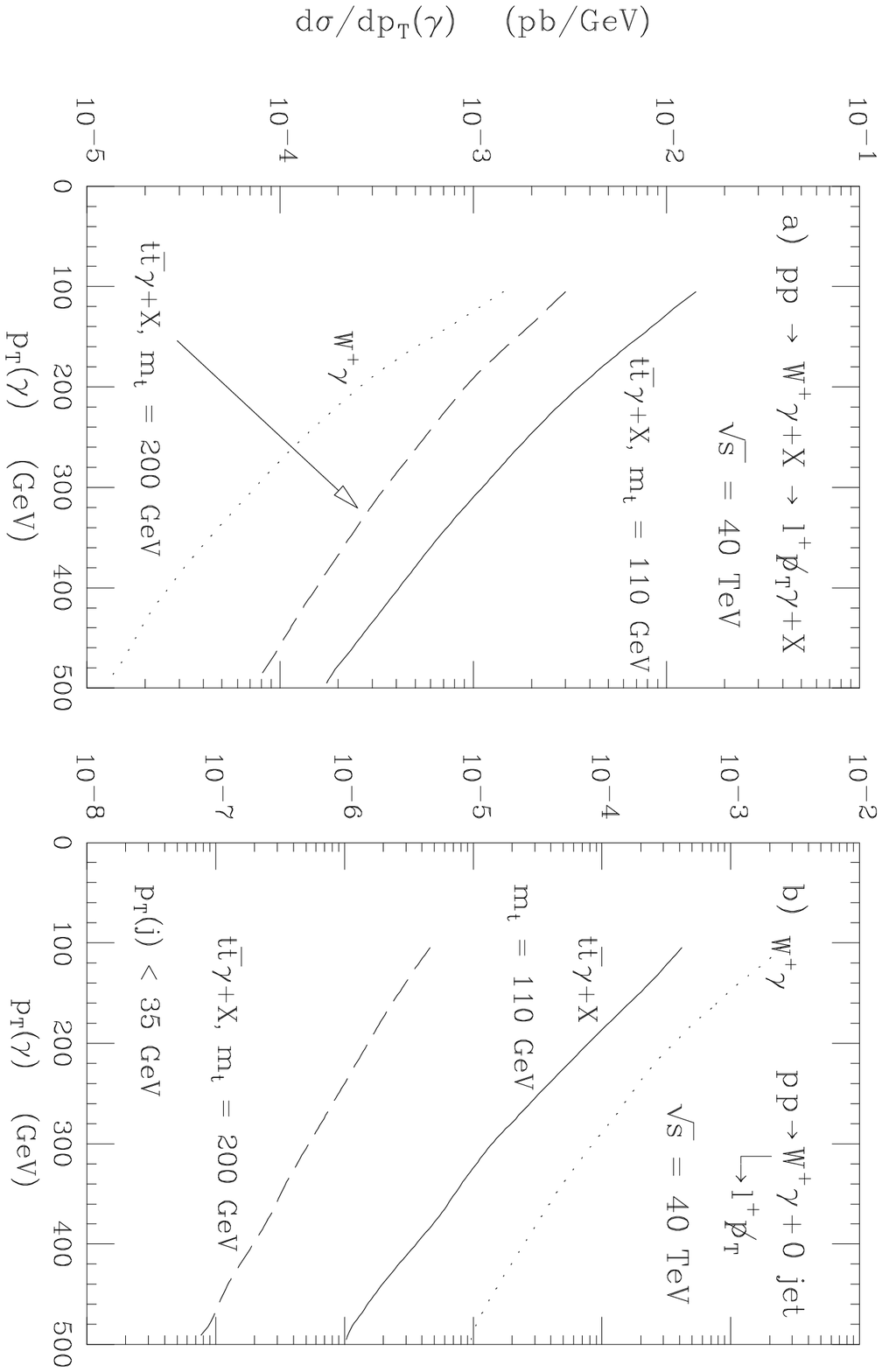}
\noindent Figure~4: a) The photon transverse momentum distribution for
$pp\rightarrow W^+\gamma+X\rightarrow\ell^+p\llap/_T\gamma+X$ at the
SSC. The solid (dashed) line shows the result for $t\bar t\gamma+X$
production for $m_t=110$~GeV (200~GeV). The dotted line gives the tree
level SM prediction of the $W^+\gamma$ signal. The cuts imposed are
summarized in Eqs.~(5) -- (7).
b) The photon transverse momentum distribution for
$pp\rightarrow W^+\gamma+0~{\rm
jet}\rightarrow\ell^+p\llap/_T\gamma+0$~jet at the SSC with a jet
transverse momentum threshold of 35~GeV. The solid (dashed) line shows the
result for the $t\bar t\gamma$ background for $m_t=110$~GeV (200~GeV).
The dotted line gives the NLO prediction of the
$p_T(\gamma)$ distribution for the $W^+\gamma+0$~jet signal in the SM.
\end{figure}
The photon bremsstrahlung cross section is approximately
40~--~65\% of the lowest order $pp\rightarrow t\bar t\gamma$ rate over
the entire $p_T(\gamma)$ range shown in Fig.~4a. The shape of the photon
transverse momentum distribution depends on the top quark mass, with
the $p_T(\gamma)$ distribution becoming harder for increasing values of
$m_t$. The dotted curve in Fig.~4a shows the lowest order prediction of the
photon $p_T$ distribution for the $W^+\gamma$ signal. The
$t\bar t\gamma$ background is seen to be much
larger than the cross section of the signal over the entire
top quark mass range studied. It is obvious from Fig.~4a that the $t\bar
t\gamma$ background will considerably reduce the sensitivity of the
reaction $pp\rightarrow W^+\gamma+X$ to non-standard $WW\gamma$ couplings.

Since the top quark decays predominantly into a $Wb$ final state, $t\bar
t\gamma$ events are characterized by a large hadronic activity which
frequently results in one or several high $p_T$ jets.
This observation suggests that the $t\bar t\gamma$ background may be
suppressed by vetoing high $p_T$ jets. Such a ``zero jet'' requirement
has been demonstrated~\cite{NLOTWO} to be very useful in reducing the size
of NLO QCD corrections in $pp\rightarrow W\gamma+X$ at SSC energies.
Present studies~\cite{SDC} suggest that for $p_T(j)<35$~GeV SSC
detectors face increasing difficulties in reconstructing jets.
Requiring that {\elevenit no} jets with $p_T(j)>35$~GeV and {\elevenit
no} second
charged lepton are observed for $|\eta(j)|,\, |\eta(\ell)|<3$, and that
the photon is isolated from the hadronic activity, one obtains the
results shown in Fig.~4b. The $t\bar t\gamma$ background is seen to be
about one order of
magnitude below the signal (dotted line) for $m_t=110$~GeV (solid line),
and approximately two orders of magnitude for $m_t=200$~GeV (dashed
line).
\vglue 0.2cm
{\elevenit\noindent 4.4 Measuring Three Vector Boson Couplings in
$qq\rightarrow qqW$ at the SSC.}
\vglue 0.1cm
In order to probe the vector boson self interactions as completely as
possible one would like to have available a large
number of basic processes and correspondingly a large number of
observables. As an example of a new process which is complementary to
weak boson pair production in the measurement of the $WW\gamma$ and
$WWZ$ vertices, the working group investigated
single $W$ production via the electroweak process $qq\rightarrow qqW$.
While $W\gamma$, $WZ$, or $W^+W^-$ production at hadron or
$e^+e^-$ colliders probe the three vector boson couplings for time-like
momenta of all interacting
electroweak bosons, the signal process measures these couplings for
space-like momentum transfer of two of the three gauge bosons.

The signal and all background cross sections were calculated using
parton level
Monte Carlo programs. For the signal the program evaluates the tree level
cross sections for the process $q_1q_2\to q_3q_4\ell\nu$, $\ell=e,\mu$ and
all relevant crossing related processes.
Representative Feynman graphs are shown in Fig.~5.
\begin{figure}[b]
\vglue 0.8in
\includegraphics{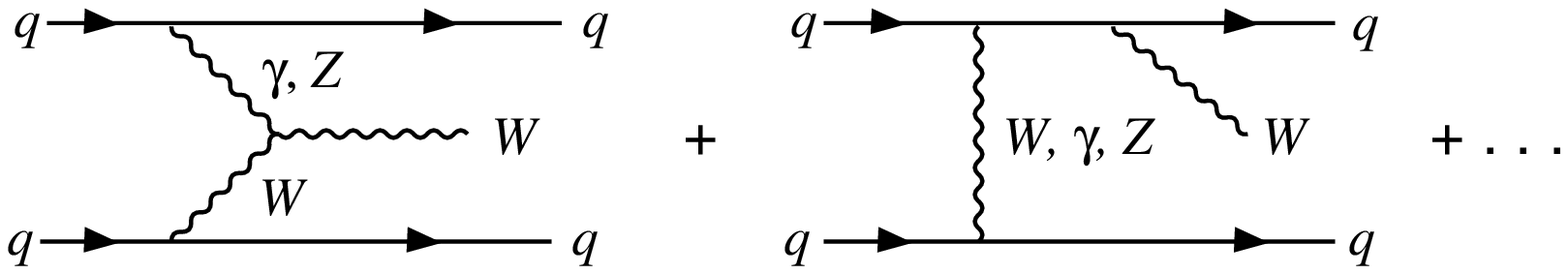}
Figure~5: Feynman graphs for the $qq\to qqW$ signal. It is via the
first graph that this process is sensitive to the three vector boson
couplings.
\end{figure}
A characteristic
feature of the $W$ signal is the presence of two energetic forward
jets. However, anticipating large backgrounds, all the features of the final
state $Wjj$ system need to be exploited for background suppression. Therefore,
the signal is only considered in the case when both final state
(anti)quarks have
transverse momenta larger than 40 GeV, allowing their identification as
hadronic jets. The potentially most dangerous background processes
are QCD $Wjj$ and $t\bar t$ production. The dominant source of forward jets in
$t\bar t$ events arises from QCD radiation, {\elevenit i.e.}\ the
additional parton
in $t\bar t j$ events, and not from the top decay products. Hence the
top background is modeled with a tree level MC program for the
process $pp \to t\bar t j\to W^+b\, W^-\bar b j$.

The three vector boson graph in $qq\rightarrow qqW$ is enhanced in the
phase space region where the two final state quark jets emerge at very
small angles. Hence events are tagged with one
very forward and one very backward jet while the lepton originating
from the $W\to \ell\nu$ decay is to be expected in the central region.
Events are triggered by a charged lepton with $p_T(\ell) > 20$~GeV,
and we require $p\llap/_T>50$~GeV as a signature
for $W$ leptonic decays. On either side of the charged lepton (with respect
to pseudorapidity) one then searches for the first hadronic jet with
$p_T(j) > 40$~GeV and $|\eta(j)|<5$,
which will be called tagging jets and represent the two spectator quarks in
our signal calculation. Leptons and jets are required to be well
separated,
$\Delta R(j,j) > 0.7$, $\Delta R(\ell,j) > 0.7$. The forward-backward
nature of the two tagging jets is then taken into account by requiring
$-5<\eta(j_1)<-2.5$ and $2.5<\eta(j_2)<5$.
Notice that this implies the existence of a central ``rapidity gap'', at
least 5~units wide in pseudorapidity, which contains the charged lepton but
no jets with $p_T>40$~GeV.

The above requirements leave a QCD $Wjj$ background which is about a factor
six larger than the remaining signal. However, the background is dominated by
$W$-bremsstrahlung off initial or final state quarks.
$W$-bremsstrahlung and the three vector boson graph lead to
drastically different lepton pseudorapidity distributions for
the signal and the background. The QCD $Wjj$ background can thus be
further suppressed by requiring $|\eta(\ell)|<1.5$ and
$\Delta\eta(\ell,j)={\rm min}(|\eta(\ell)-\eta(j_1)|,
|\eta(\ell)-\eta(j_2)|) > 2.5$. An additional strong background
rejection is achieved by exploiting the very
large dijet invariant masses, $m(jj)$, which are typical for the vector boson
fusion process. A cut of $m(jj)>3$~TeV,
imposed on the two tagging jets, reduces the background well below signal
level
($\sigma_{\rm SIG} = 450$~fb vs. $\sigma_{\rm QCD} = 136$~fb for the QCD $Wjj$
background and $\sigma_{t\bar tj} = 38$~fb for $m_t=140$~GeV).

The cuts discussed above single out the phase space region
in which the electroweak fusion process dominates and hence one expects a
pronounced sensitivity to deviations in the $WW\gamma$ and $WWZ$
couplings from the SM prediction. Anomalous coupling
effects are enhanced at large momentum transfer and, hence, for large
transverse momenta of the produced $W$-boson. The effect is demonstrated in
Fig.~6.
\begin{figure}[t]
\vglue3.in
\includegraphics{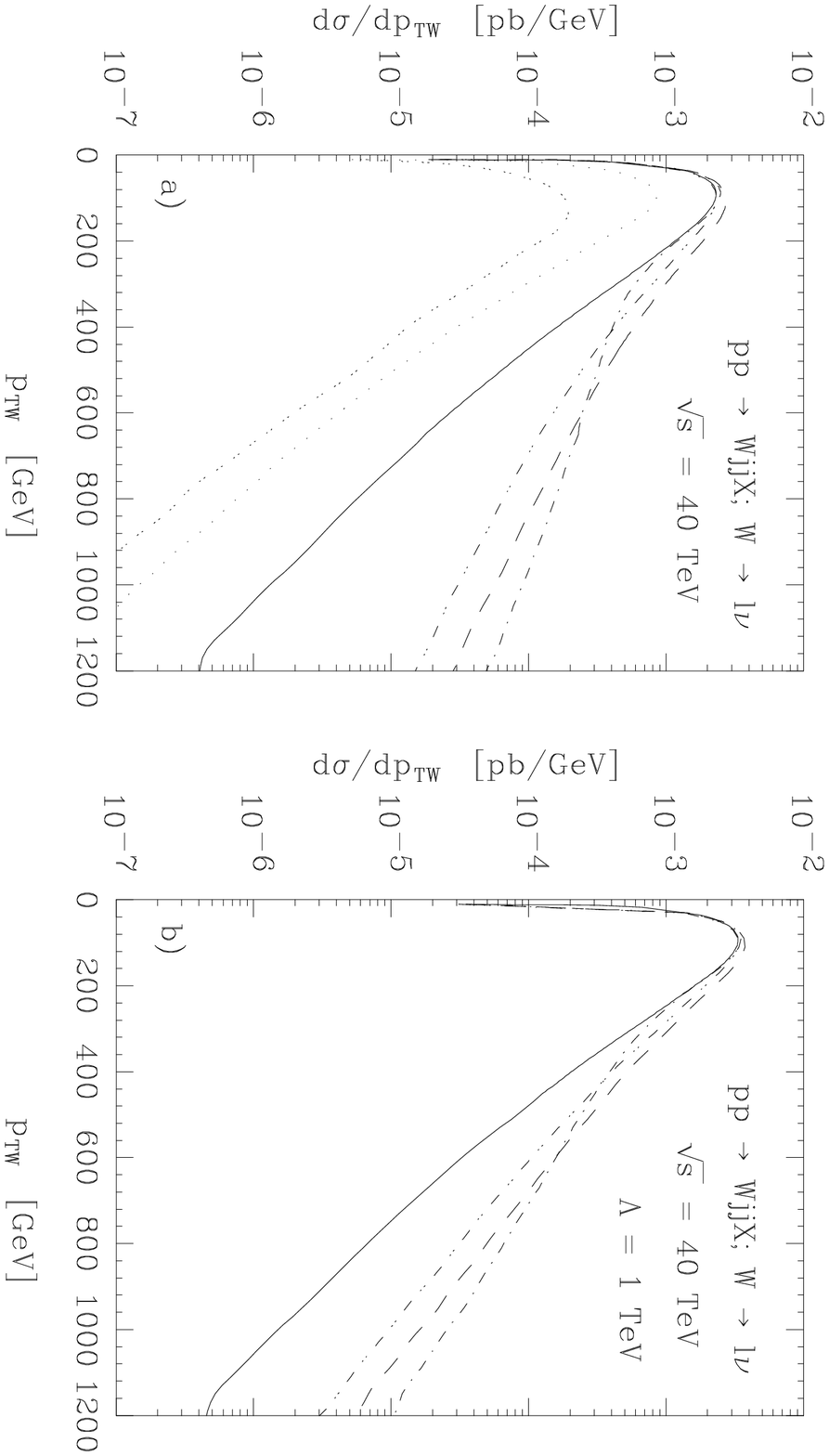}
Figure~6: Transverse momentum distribution of the produced $W$-boson in
$Wjj$ events at the SSC. a) Individual distributions for the SM
signal (solid line), the QCD $Wjj$ background (dotted line) and the $t\bar
tj$ background for $m_t=140$~GeV (double dotted line). The upper three
curves correspond to three choices of anomalous couplings: $\kappa^\gamma=
\kappa^Z=1.2$ (dashed line) $\lambda^\gamma=\lambda^Z=0.1$ (dash-dotted curve)
and $g_1^Z=1.2$ (dash-double dotted line). b) The two background
distributions have been added to the four signal curves. In addition the
effect of a form-factor is shown for a scale
$\Lambda=1$~TeV. The cuts imposed are described in the text.
\end{figure}
While the $p_{TW}$ distributions show similar shapes for the SM
signal and the QCD and top quark backgrounds, a strong enhancement at large
transverse momenta arises from anomalous couplings. For these three curves all
other anomalous couplings are set to zero.
Here the standard parameterization of the anomalous couplings in terms
of $C$ and $P$ conserving anomalous couplings $g_1^V,\; \kappa^V$ and
$\lambda^V$, $V=\gamma,\, Z$ is used. These couplings can be defined by
the effective Lagrangian
\begin{equation}
i{\cal L}_{eff}^{WWV} = g_{WWV}\left( g_1^V(
W^{\dagger}_{\mu\nu}W^{\mu}-W^{\dagger\, \mu}W_{\mu\nu})V^{\nu} +
\kappa^V\,  W^{\dagger}_{\mu}W_{\nu}V^{\mu\nu} + {\lambda^V\over m_W^2}\,
W^{\dagger}_{\rho\mu}{W^{\mu}}_{\nu}V^{\nu\rho}\right) , \label{LeffWWV}
\end{equation}
where the overall coupling constants are defined as $g_{WW\gamma}=e$ and
$g_{WWZ}= e \cot\theta_W$. Within the SM, at tree level,
$g_1^Z = g_1^\gamma = \kappa^Z = \kappa^\gamma = 1$, and $\lambda^Z =
\lambda^\gamma = 0$. $g_1^\gamma$ is just the electric charge of the $W$ and
hence fixed to 1 by electromagnetic gauge invariance.

{}From Fig.~6b it can be seen
that form factor effects are important in $qq\rightarrow qqW$. The
results of a more quantitative analysis of the sensitivity of $Wjj$
production to anomalous $WWV$ couplings are described in
Ref.~\cite{UD}. The $2\sigma$ limits on $\kappa^V-1$, $\lambda^V$ and
$\Delta g_1^Z=g_1^Z-1$ are found to be in the $0.03\dots 0.1$ range.
Comparing these results with the limits obtained from di-boson
production, one finds that the process
$qq\rightarrow qqW$ is significantly more sensitive to $\Delta\kappa^V$
and $\Delta g_1^Z$ than $W\gamma$ and $WZ$ production for cutoff scales
$\Lambda$ in the low TeV range. In general, the pair production process
is affected more by details of the
form-factors. This emphasizes the need to measure $Wjj$ production in
addition to di-boson production if full information on the $WWV$
couplings is to be gained.
\newpage
{\elevenbf\noindent 5. Summary}
\vglue 0.2cm
The large statistics accumulated in the 1992-94 Tevatron runs will result in
significantly
improved measurements of the SM parameters. The energy distribution of
the $W$ decay electron offers an attractive alternative way to measure
$m_W$. The large number of
$W+1$~jet events will make it possible to measure $W$ plus charm production,
which could help constraining the $s$-quark distribution function. Vector
boson self-interactions are expected to be probed with 10~--~60\%
accuracy with the new Tevatron data. The lepton photon rapidity
difference distribution will give easy access to the radiation zero
predicted by the SM for $W\gamma$ production. $t\bar t\gamma$
production was found to be an important background to $W\gamma$
production at the SSC. However, it can be easily eliminated by a jet veto.
Single
$W$ production via electroweak interactions at the SSC provides a
measurement of the $WWV$ vertices which is complementary to that in
di-boson production.
\vglue 0.3cm
{\elevenbf\noindent 6. References}
\vglue 0.2cm

\end{document}